\documentstyle[aps,epsfig,floats]{revtex}

\newcommand\beq{\begin{equation}}
\newcommand\eeq{\end{equation}}
\newcommand\bea{\begin{eqnarray}}
\newcommand\eea{\end{eqnarray}}
\def\non{\nonumber}
\def\bib{\bibitem}

\def\v{\vert}

\begin{document}

\draft

\textheight=23.8cm

\title{\Large \bf Quantum bound states for a derivative nonlinear 
Schr\"odinger model \\
and number theory}
\author{\bf B. Basu-Mallick$^1$ \footnote{E-mail: biru@theory.saha.ernet.in}, 
Tanaya Bhattacharyya$^1$ \footnote{E-mail: tanaya@theory.saha.ernet.in}, 
and Diptiman Sen$^2$ \footnote{E-mail: diptiman@cts.iisc.ernet.in}}
\address{$^1$Theory Group, Saha Institute of Nuclear Physics,
1/AF Bidhan Nagar, Kolkata 700 064, India \\
$^2$Centre for Theoretical Studies, Indian Institute of Science,
Bangalore 560012, India}

\maketitle

\begin{abstract}
A derivative nonlinear Schr\"odinger model is shown to support localized 
$N$-body bound states for several ranges (called bands) of the coupling 
constant $\eta$. The ranges of $\eta$ within each band can be completely 
determined using number theoretic concepts such as Farey sequences and 
continued fractions. For $N \ge 3$, the $N$-body bound states can have both 
positive and negative momentum. For $\eta > 0$, bound states with positive 
momentum have positive binding energy, while states with negative momentum 
have negative binding energy.
\end{abstract}
\vskip .3 true cm


\vskip 1 true cm

Integrable quantum models in 1+1 dimensions which support bound states have 
been studied extensively for many years 
\cite{thacker,faddeev,skylanin,wadati1,bhat,shnirman,kundu,basu}. 
For an integrable Hamiltonian, the coordinate Bethe ansatz can yield the exact
eigenfunctions. If an eigenfunction decays exponentially fast when any of the 
interparticle distances tends towards infinity (keeping the center of mass 
coordinate fixed), we call such a localized 
square-integrable eigenfunction a bound state. Bound states of quantum 
integrable models are usually found to have positive binding energy 
\cite{thacker,faddeev,skylanin,wadati1,bhat}.

In this paper, we will study the quantum bound 
states of an integrable derivative nonlinear Schr\"odinger 
(DNLS) model \cite{shnirman,kundu}. Classical and quantum versions of
the DNLS model have found applications in many areas of physics like 
circularly polarized nonlinear Alfven waves in a plasma \cite{wadati2},
quantum properties of solitons in optical fibers \cite{kodama}, and 
some chiral Luttinger liquids \cite{aglietti}. The classical DNLS model is 
known to have solitons with momenta in only one direction \cite{chen,min}. 

Using the coordinate Bethe ansatz, it had been found earlier that 
quantum $N$-body bound states exist for the DNLS model provided that the 
interaction parameter $\eta$ (defined in Eq. (\ref{ham2}) below) lies in the 
range $0 < \v \eta \v < \tan (\pi /N)$. It was also observed that, similar to 
the classical case, such $N$-body bound states can have only positive values 
of $P/\eta$, where $P$ is the momentum \cite{shnirman}. However, it was found 
recently that bound states can exist with
$P/\eta <0$ provided that $\tan (\pi /N) < \v \eta \v < \tan [\pi /(N-1)]$, 
and that these states have {\it negative} binding energy \cite{basu}. This
naturally leads one to ask: are there other ranges of values of $\eta$ for
which quantum bound states exist, and, if they exist, what are their momenta 
and binding energies?

In this paper, we will solve the problem of determining the {\it complete} 
ranges of values of $\eta$ for which quantum $N$-body bound states exist in 
the DNLS model, for {\it all} values of $N$. 
After presenting the conditions which are required for a 
quantum $N$-body bound state to exist, we will use the idea
of Farey sequences in number theory to show that there are certain ranges of 
$\eta$, called bands, in which bound states exist. We find that the bound 
states appearing within each band can have both positive and negative
values of $P/\eta$; these have positive and negative binding energies 
respectively. We will then use another concept from number theory, that 
of continued fractions, to address the inverse problem of finding the values 
of $N$ for which $N$-body bound states exist for a given value of $\eta$. 

For $N$ particles, the Hamiltonian of the DNLS model is given by
\beq
H_N ~=~  -\hbar^2 ~\sum_{j=1}^N ~\frac{\partial^2}{\partial x_j^2} ~+~ i2 
\hbar^2 \eta ~\sum_{l<m} ~\delta(x_l - x_m )~ \Big( 
\frac{\partial}{\partial x_l} + \frac{\partial} {\partial x_m} \Big) ~,
\label{ham2}
\eeq
where we have set the particle mass $m=1/2$.
$H_N$ commutes with the momentum operator
$P_N = -i\hbar \sum_{j=1}^N \partial /\partial x_j$. We note that $H_N$ 
remains invariant while $P_N$ changes sign if we change the sign of $\eta$ and 
transform all the $x_i \rightarrow - x_i$ at the same time; we call this 
the parity transformation. Hence it is sufficient to study the model for
one particular sign of $\eta$, say, $\eta >0$. The eigenfunctions for 
$\eta <0$ can then be obtained by changing $x_i \rightarrow - x_i$.

Next, the coordinate space $R^N \equiv \{ x_1, x_2, \cdots x_N \}$ is divided 
into various $N$-dimensional sectors defined through inequalities like 
$x_{\omega(1)}< x_{\omega(2)}< \cdots < x_{\omega(N)}$, where $\omega(1), 
\omega(2), $ $\cdots , \omega(N)$ represents a permutation of the integers
$1,2, \cdots ,N$. Given the wave function in the fundamental sector,
defined as $x_1 < x_2 < \cdots < x_N$, the wave functions in all the other
sectors can be found by Bose symmetry. In the fundamental sector, the 
Bethe ansatz wave function takes the form 
\beq
\psi = \sum_\omega ~C_\omega ~\exp ~\{ i (k_{\omega(1)} 
x_1 + \cdots + k_{\omega(N)} x_N ) \},
\label{psi}
\eeq
where $k_n$'s are all distinct wave numbers, the sum is over all permutations
$\omega$ of the integers $1,2, \cdots ,N$, and $C_\omega$ are appropriate 
coefficients. The momentum and energy of this eigenfunction are given by
\beq
P ~=~ \hbar \sum_{j=1}^N k_j ~, \quad {\rm and} \quad E ~=~ \hbar^2 
\sum_{j=1}^N k_j^2 ~.
\label{momen1}
\eeq

In Ref. \cite{shnirman}, it was shown that a localized bound state has
a wave function consisting of only one plane wave in each sector. Namely,
in the fundamental sector, the coefficients $C_\omega$ in (\ref{psi})
vanish for all $\omega$'s except for the identity permutation.
Further, the momenta $k_n$'s satisfy the following conditions:
\beq
k_n - k_{n+1} ~+~ i ~\eta ~(k_n+ k_{n+1}) ~=~ 0 ~,
\label{cond1}
\eeq
for $n=1,2,\cdots ,N-1$,
\beq
\sum_{j=1}^N q_j ~=~ 0 ~,
\label{cond2}
\eeq
where $q_j$ denotes the imaginary part of $k_j$, and
\beq
q_1< 0 ~, ~~~~q_1+q_2 < 0 ~, ~~\cdots ~~, ~\sum_{j=1}^{N-1} ~q_j < 0 ~. 
\label{cond3}
\eeq
Eqs. (\ref{cond2}-\ref{cond3}) imply that the wave function $\psi$ is 
square-integrable if one holds the center of mass coordinate $X =\sum_i 
x_i /N$ fixed, and integrates over the relative coordinates $y_r = x_{r+1} - 
x_r$, where $r=1,2, \cdots , N-1$. In the fundamental sector, the integrals 
over the $y_r$'s all run from $0$ to $\infty$, and they are independent of 
each other. Using Eq. (\ref{cond2}-\ref{cond3}), one can express the 
probability density as
\beq
| \psi |^2 ~\sim ~ \exp [ ~\sum_{r=1}^{N-1} ~( \sum_{j=1}^r q_j ) ~y_r ~]  \, ,
\label{psi2}
\eeq
which is independent of $X$. Due to the conditions in (\ref{cond3}), 
integration of this probability density over the $y_r$'s gives a finite result.

The conditions (\ref{cond1}) and (\ref{cond2}) imply that the
$k_n$'s must be complex numbers of the form
\beq
k_n ~=~ \chi ~e^{-i(N+1-2n)\phi} ~,
\label{kn}
\eeq
where $\chi$ is a real parameter, and $\phi \equiv \tan^{-1} 
\eta$. Since $0 < \v \eta \v < \infty$, we assume that $0 < \v \phi \v < 
\pi /2$. Using Eqs. (\ref{momen1}) and (\ref{kn}), the momentum 
and energy of this state are found to be
\beq
P ~=~ \hbar \chi ~\frac{\sin (N\phi)}{\sin \phi} ~, \quad {\rm and} \quad
E ~=~ \frac{\hbar^2 \chi^2 \sin(2N \phi)}{\sin(2\phi)} ~.
\label{momen2}
\eeq
If we define the mass of this state by the relation $E=P^2 /(2M)$, we find that
\beq
M ~=~ \frac{\tan (N\phi)}{2 \tan (\phi)} ~.
\label{mass}
\eeq

We now impose the conditions (\ref{cond3}) on the $k_n$'s. We find that
all the following inequalities must be satisfied,
\beq
\chi ~\frac{\sin (l \phi)}{\sin \phi} ~\sin [(N-l) \phi] ~>~ 0 
\label{cond4}
\eeq
for $l=1,2,\cdots , N-1$. 
For $N=2$, (\ref{cond4}) is satisfied when $\phi$ lies in the range
$0 < \phi < \pi /2$ ($- \pi /2 < \phi < 0$) if $\chi >0$ ($\chi < 0$). Thus 
any nonzero value of $\phi$ allows a $2$-body bound state. From (\ref{momen2}), 
we see that the ratio $P/\phi >0$.
 
We now consider the more interesting case with $N \geq 3$. Due to the parity 
symmetry of (\ref {ham2}), we will henceforth assume that $\phi > 0$ (i.e. 
$\eta >0$). Eq. (\ref{cond4}) can then be rewritten as
\beq
\chi ~\sin (l\phi) ~\sin [(N-l) \phi] ~>~ 0 
\label{cond5}
\eeq
for $l=1,2,\cdots ,N-1$. For $\chi > 0$, Eq. (\ref{cond5}) implies
\beq
\cos [ (N-2l) \phi ] ~>~ \cos (N\phi) 
\label{chi1}
\eeq
for $l=1,2,\cdots ,N-1$. Let us now consider a value of $\phi$ of the form
\beq
\phi_{N,n} ~\equiv ~\frac{\pi n}{N} ~,
\label{phinn}
\eeq
where $n$ is an integer satisfying $1 \le n < N/2$. If $n$ is odd,
$\cos (N \phi_{N,n}) = -1$. We then find that all the
inequalities in (\ref{chi1}) are satisfied provided that $N$ and $n$ are 
relatively prime, i.e., if the greatest common divisor of $N$ and $n$ is 1.
Similarly, for $\chi < 0$, Eq. (\ref{cond5}) takes the form
\beq
\cos [ (N-2l) \phi ] ~<~ \cos (N\phi) 
\label{chi2}
\eeq
for $l=1,2,\cdots ,N-1$. We find that all these inequalities 
are satisfied if $n$ is even, and $N$ and $n$ are relatively prime.

In short, all the inequalities in (\ref{cond5}) are satisfied for $\phi = 
\phi_{N,n}$, if and only if $N$ and $n$ are relatively prime (with $n$ odd 
for $\chi >0$, and $n$ even for $\chi < 0$). By continuity, it follows that 
all the inequalities will hold in a neighborhood of $\phi_{N,n}$ extending 
from a value $\phi_{N,n,-}$ to a value $\phi_{N,n,+}$.
The region $\phi_{N,n,-} < \phi < \phi_{N,n,+}$ will be called the band 
$B_{N,n}$. 

For a given value of $N$, the number of bands in which bound states exist
is equal to the number of integers $n$ which are relatively prime to $N$ and
satisfy $1 \le n < N/2$. This is equal to half the number of integers which 
are relatively prime to $N$ and satisfy $1 \le n < N$. The latter number is 
called Euler's $\phi$-function $\Phi (N)$ \cite{niven}. The number of bands 
is therefore equal to $\Phi (N) /2$ for $\eta > 0$.

We now have to determine the end points $\phi_{N,n,-}$ and $\phi_{N,n,+}$
of the band $B_{N,n}$. One or more of the inequalities in (\ref{cond5}) will
be violated at the end points $\phi_{N,n,\pm}$ if 
\beq
\phi_{N,n,\pm} ~=~ \frac{\pi j_{\pm}}{l_{\pm}} ~,
\label{jl1}
\eeq
where $j_{\pm}$ and $l_{\pm}$ are integers satisfying 
\beq
1 ~\le ~ l_{\pm} ~<~ N ~, \quad {\rm and} \quad j_{\pm} ~< ~\frac{l_{\pm}}{2}
\label{jl2}
\eeq
(since $\phi < \pi /2$). Thus the end points of the band $B_{N,n}$ are given 
by two rational numbers of the form $j_{\pm}/l_{\pm}$ which lie {\it closest} 
to (and on either side of) the point $\phi_{N,n} /\pi = n/N$. These can be 
found using the idea of Farey sequences \cite{niven}.

For a positive integer $N$, the Farey sequence $F_N$ is defined to be the set
of all the fractions $a/b$ in increasing order such that (i) $0 \le a \le b 
\le N$, and (ii) $a$ and $b$ are relatively prime. 
For $N \ge 2$, if $n/N$ is a fraction appearing somewhere in the sequence 
$F_N$, then it is known that the fractions $a_1/b_1$ and $a_2/b_2$ appearing
immediately to the left and to the right respectively of $n/N$ satisfy
\bea
a_1 ~,~ a_2 ~\le ~n ~, & & \quad {\rm and} \quad a_1 ~+~ a_2 ~=~ n ~, \non \\
b_1 ~,~ b_2 ~<~ N ~, & & \quad {\rm and} \quad b_1 ~+~ b_2 ~=~ N ~, \non \\
n b_1 ~-~ N a_1 = 1 ~, & & \quad {\rm and} \quad n b_2 ~-~ N a_2 = - ~1 ~,
\label{fs2}
\eea
and $n, ~b_1, ~b_2$ are relatively prime to $N$.

\begin{figure}[htb]
\vspace*{-1.4cm}
\begin{center}
\epsfig{figure=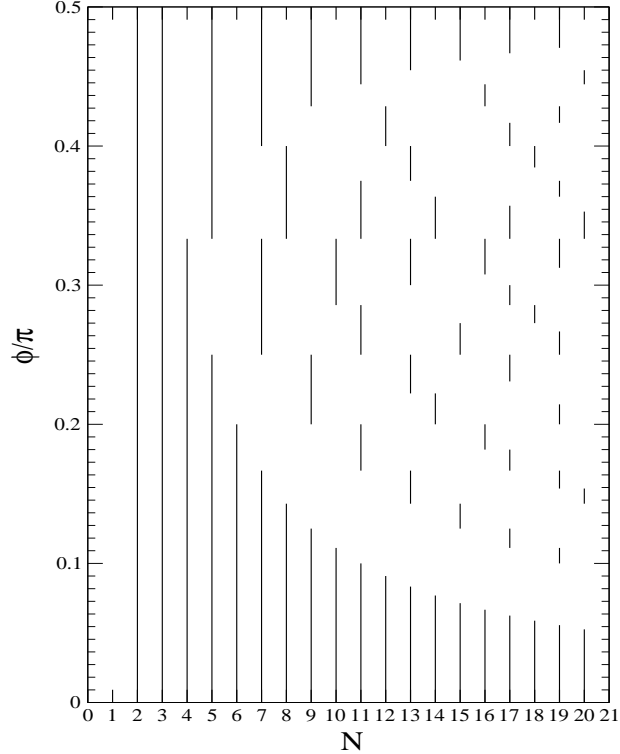,height=12cm,width=10cm}
\end{center}
\caption{The values of $\phi /\pi$ for which $N$-body bound states exist 
for various values of $N$.}
\end{figure}

Using Eqs. (\ref{jl1}) and (\ref {jl2}), we now see that the end points of the
band $B_{N,n}$ are given by
\beq
\phi_{N,n,-} ~=~ \frac{\pi a_1}{b_1} ~, \quad {\rm and} \quad 
\phi_{N,n,+} ~=~ \frac{\pi a_2}{b_2} ~,
\label{ends}
\eeq
where $a_1/b_1$ and $a_2/b_2$ are the fractions lying to the left and right 
of $n/N$ in the Farey sequence $F_N$.

For $N \ge 3$, the lowest band is given by $n=1$; by using Eq. (\ref{fs2}),
the range of this band is obtained as $0 < \phi/\pi < 1/(N-1)$.
For higher values of $n$, the end points of the 
band $B_{N,n}$ (i.e., the integers $a_i$ and $b_i$) can be determined 
numerically by using the properties given in Eq. (\ref{fs2}). Fig. 1 shows 
the ranges of values of $\phi$ for which bound states exist for $N=2$ to 20. 

Eq. (\ref{fs2}) implies that the width of the
right side of the band $B_{N,n}$ from $\phi_{N,n}$ to $\phi_{N,n,+}$ is
$\pi /(N b_2)$, while the width of the left side from $\phi_{N,n,-}$ to 
$\phi_{N,n}$ is $\pi /(N b_1)$. For later use, we note that each of
these widths is larger than $\pi /N^2$, since $b_1, ~b_2 < N$.

We now calculate the momentum and binding energy for the 
$N$-body bound states in a particular band $B_{N,n}$ using Eq. 
(\ref{momen2}). The form of the end points given in Eq. (\ref{ends}) shows
that $\sin (N \phi) = 0$ at only one point in the band $B_{N,n}$, namely, at 
$\phi = \phi_{N,n}$. In the right part of the band (i.e., from $\phi_{N,n}$
to $\phi_{N,n,+}$), the sign of $\sin (N\phi)$ is $(-1)^n$. In the left part 
of the band (i.e., from $\phi_{N,n,-} $ to $\phi_{N,n}$), the sign of $\sin 
(N\phi)$ is $(-1)^{n+1}$. Since $\chi$ has the same sign as $(-1)^{n+1}$, the 
momentum given in Eq. (\ref{momen2}) is positive in the left part of the band,
negative in the right part of the band, and zero at $\phi = \phi_{N,n}$. 

To calculate the binding energy, we consider a reference state in which the 
momentum $P$ of the $N$-body bound state is equally distributed among $N$ 
single-particle scattering states. From Eqs. (\ref{momen1}) and (\ref{momen2}),
the wave number associated with each of these single-particle states is found 
to be $k_0 = \chi \sin(N \phi) /(N \sin \phi)$. The total 
energy for the $N$ single-particle scattering state is therefore given by
\bea
E_s ~=~ \hbar^2 N k_0^2 ~=~ \frac{\hbar^2 \chi^2 \sin^2 (N \phi)}{N\sin^2 
\phi} ~.
\label{es}
\eea
Subtracting $E$ in (\ref{momen2}) from $E_s$ in (\ref{es}), we obtain the 
binding energy of the $N$-body bound state as 
\beq
E_B (\phi, N) ~=~ \frac{\hbar^2 \chi^2\sin (N \phi)}{\sin \phi} 
\Big\{\frac{\sin (N \phi)}{N\sin \phi} -\frac{\cos (N \phi)}{\cos \phi} \Big\}.
\label{eb1}
\eeq

\begin{figure}[htb]
\vspace*{-3cm}
\begin{center}
\epsfig{figure=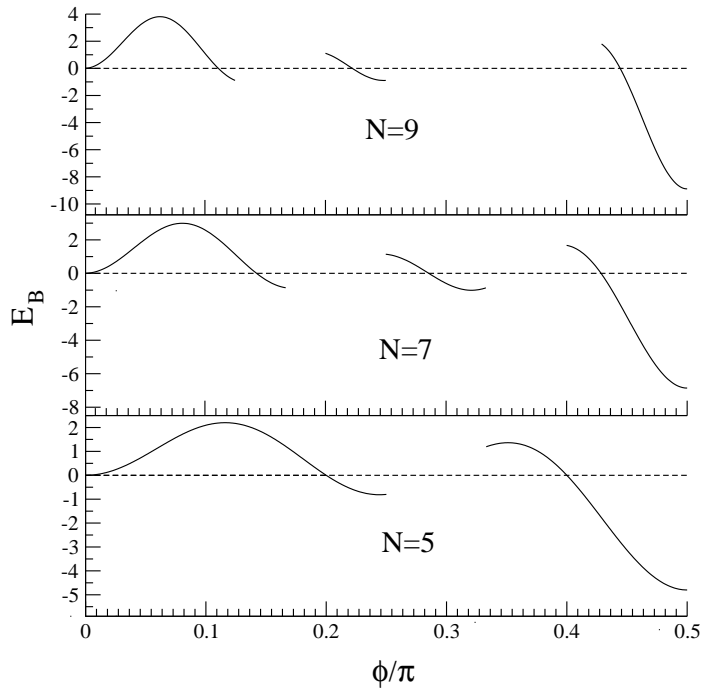,height=14cm,width=10cm}
\end{center}
\vspace*{-1cm}
\caption{The binding energy $E_B$ of the $N$-body bound state as a function of
$\phi /\pi$ for three different values of $N$.}
\end{figure}

Substituting $N=2$ in Eq. (\ref{eb1}), we obtain $E_B(\phi, 2) =2\hbar^2 \chi^2
\sin^2 \phi$. Thus $E_B (\phi,2) >0$ for any nonzero value of $\phi$. Let us 
now consider the case $N \ge 3$. We can rewrite Eq. (\ref{eb1}) in the form
\bea
E_B (\phi ,N) &=& \frac{\hbar^2 \chi \sin (N \phi)}{N \sin^2 \phi \cos \phi} ~
f (\phi , N) ~, \non \\
f (\phi , N) &=& \chi ~[\sin (N \phi) \cos \phi ~-~ N \cos (N \phi) \sin 
\phi ].
\label{eb2}
\eea
On adding up all the 
inequalities given in (\ref{chi1}) or (\ref{chi2}), and using the identity
$\sum_{l=1}^{N-1} \cos [(N-2l) \phi] = \sin [(N-1) \phi] /\sin \phi$, we find
that $f (\phi , N)$ is positive in all the bands $B_{N,n}$ for all values of 
$N$ and $n$. Hence, $E_B$ given in (\ref{eb2}) has the same sign as $\chi \sin
(N \phi)$. Following arguments similar to that of the momentum, we find that 
the binding energy is positive in the left part of each band, negative in the 
right part, and zero at the point $\phi = \phi_{N,n}$.

We thus see that for $\phi >0$, the momentum and the binding energy are both 
positive in the left part of each band, and they are both negative in the right
part. [If $\phi < 0$, we can similarly show that bound states with positive 
(negative) values of $P/\phi$ have positive (negative) binding energy]. In Fig.
2, we show the binding energy $E_B$ as a function of $\phi /\pi$ for three 
different values of $N$. (We have set $\hbar^2 \chi^2 =1$ in the figure). We 
see that $E_B$ is indeed positive (negative) in the left (right) part of each 
band.

We will now use the technique of continued fractions to 
study the inverse problem of determining the values of $N$ for which $N$-body 
bound states exist for a given value of $\phi$. Any positive real
number $x$ has a simple continued fraction expansion of the form \cite{niven}
\beq
x ~=~ n_0 ~+~ \frac{1}{n_1 ~+~ \frac{1}{n_2 ~+~ \cdots}} ~,
\eeq
where the $n_i$'s are integers satisfying $n_0 \ge 0$, and $n_i \ge 1$ for $i
\ge 1$. The expansion ends at a finite stage with a last integer $n_k$ if $x$ 
is rational, and does not end if $x$ is irrational.
Given a number $x$, the integers $n_i$ can be found as follows. We define
$x_0 = x$. Then $n_0 = [x_0]$, where $[y]$ denotes the integer part of a
non-negative number $y$. We then recursively define $x_{i+1} = 1/(x_i - n_i)$,
and obtain $n_{i+1} = [x_{i+1}]$ for $i =0,1,2,\cdots$. If we stop at the 
$k^{\rm th}$ stage, we obtain a rational number $r_k =<n_0,n_1,n_2,\cdots ,
n_k>$ which is an approximation to the number $x$. If we write $r_k = p_k/q_k$,
where $p_k$ and $q_k$ are relatively prime, then it is known that
\beq
|~ x ~-~ \frac{p_k}{q_k} ~| ~<~ \frac{1}{q_k^2} ~, 
\label{cf}
\eeq
for all values of $k \ge 1$ \cite{niven}.

Now suppose that we know the expansion 
\beq
\frac{\phi}{\pi} ~=~ <0,n_1,n_2,\cdots > ~.
\label{cfphi}
\eeq
If we stop at the $k^{\rm th}$ stage in this expansion, we obtain $p_k /q_k 
= <0,n_1,n_2,\cdots , n_k>$. Eq. (\ref{cf}) then implies that 
\beq
|~ \frac{\phi}{\pi} ~-~ \frac{p_k}{q_k} ~| ~<~ \frac{1}{q_k^2} ~.
\label{cfbound}
\eeq
We now recall the comment that both the right and the left part of
the band $B_{q_k,p_k}$ have widths which are larger than $1/q_k^2$.
Hence Eq. (\ref{cfbound}) implies that $\phi /\pi$ must lie within the 
band $B_{q_k,p_k}$. We have thus found a value of $N=q_k$ for which
an $N$-body bound state exists for the given value of $\phi$.
We can generate several such values of $N$ by stopping at different stages
$k$ in the expansion given in (\ref{cfphi}). If $\phi /\pi$ is rational, the
continued fraction expansion stops at a finite stage, so we only obtain
a finite number of values of $N$ in this way. This can also be seen directly 
from Eq. (\ref{cond5}). If $\phi /\pi = p/q$ is rational, then at least one of 
the inequalities in (\ref{cond5}) will be violated if $N > q$. 
We thus conclude that if $\phi /\pi$ is rational, there is only a finite
number of values of $N$ for which a $N$-body bound state exists. 
If $\phi /\pi$ is irrational, then the expansion in (\ref{cfphi}) does not 
end, and we can use the procedure described above to find an infinite number 
of possible values of $N$ for which a $N$-body bound state exists.

To conclude, we have used the ideas of Farey sequences and continued 
fractions to determine all the allowed ranges (bands) of $\eta$ for which 
quantum $N$-body bound states exist in the DNLS model. For $N \ge 3$, we find
that the $N$-body bound states can have both positive and negative momentum. 
Bound states with positive (negative) values of $P/\eta$ have positive 
(negative) binding energy. Our work brings the analysis of the quantum bound 
states in the DNLS model to the same level of completion as that of the usual 
nonlinear Schr\"odinger model (where bound states are known to exist for all 
negative values of the coupling constant and all values of $N \ge 2$).

Bound states with negative binding energy are unusual in the field of
integrable quantum models. However, such states are known to exist in other 
areas of quantum physics, such as antibonding states in molecules (see 
\cite{harrison} for instance). The negative binding energy states that we
have found in the DNLS are stable because the model is integrable. Presumably,
these states would decay if one were to add terms to the Hamiltonian which
destroy the integrability; any real system would probably have such terms
anyway, so it is not clear at the moment if such states can be observed 
experimentally.

The quantum bound states which exist in the lowest band and have positive 
binding energy can be related in several ways to the solitons which appear in
the classical version of the DNLS model which is integrable.
A general method for relating quantum bound states
and classical solitons is described in Ref. \cite{wadati1}.
For the case of DNLS model, the classical 
solitons are localized solutions of the equation 
\beq
i \hbar ~\frac{\partial \psi}{\partial t} ~=~ - \hbar^2 ~\frac{\partial^2 
\psi}{\partial x^2} ~+~ i 4 \hbar \eta ~\psi^* \psi ~\frac{\partial 
\psi}{\partial x} ~,
\eeq
with $\int_{-\infty}^{\infty} dx \psi^* \psi = \hbar N$. These solitons are
known to exist only if $0 < \v \eta \v < \pi /N$ \cite{shnirman,min}, and they
can be obtained by taking the $\hbar \rightarrow 0 ~,~ N\rightarrow \infty$ 
limit of the quantum bound states \cite{shnirman}. 
Another way of relating the classical 
solitons and quantum bound states of DNLS model 
 is indicated in the first paper in 
Ref. \cite{aglietti}. There it is argued that the classical soliton
mass, with one-loop quantum corrections, is given by
\beq
M_{\rm cl} ~=~ \frac{1}{2} ~[ ~N ~+~ \frac{\eta^2}{3} ~(N^3 - N) ~] ~.
\eeq
Comparing this to the mass of the $N$-body bound state in Eq. (\ref{mass}), we
see that the two agree up to order $\eta^2$ for small $\eta$. 
An interesting problem for future study may be to see if there is a classical 
version of the bound states in the higher bands which we have found in this 
paper. Since the ranges of values of $\eta$ for which these bound states
exist depend sensitively on $N$, going to the limit $N \rightarrow \infty$ 
with a fixed value of $\eta$ may turn out to be a rather subtle problem.

\end{document}